\documentclass[epj]{svjour}
\usepackage{graphics}
\usepackage{graphicx}
\usepackage[latin1]{inputenc}
\usepackage[centerlast,sc,small]{subfigure}
\usepackage{bm}
\newcommand{\dd}{\mathrm{d}}
\newcommand{\bb}{\begin{eqnarray}}
\newcommand{\ee}{\end{eqnarray}}

\newcommand{\nn}{\nonumber}
\newcommand{\kb}{k_\mathrm{B}}

\begin{document}
\title{Transverse magnetization and transient oscillations in the quantum tunneling of molecular magnets}
\dedication{To Yukiko, who never questioned the beauty of nanomagnetism}
\author{Maxime Clusel\inst{1}\thanks{\emph{clusel@ill.eu}}
\and Timothy Ziman \inst{1,2}
\thanks{\emph{ziman@ill.eu}}%
}                     
%
%
\institute{Institut Laue-Langevin, 6 rue Jules Horowitz, B.P. 156X, F-38042 Grenoble Cedex, France,  \and LPM2C- CNRS, 25 avenue des Martyrs,  B.P. 166, F-38042 Grenoble Cedex, France.}
\titlerunning{Transverse magnetization and transcient oscillations in molecular magnets}
\authorrunning{M.Clusel \textit{et al.}.}
\date{\today}
%
\abstract{We calculate the response of a molecular magnet subject to a time-varying
magnetic field and coupled to a heat bath. We propose that observations of 
calculated oscillations transverse to the field direction may be an effective way
of demonstrating quantum tunneling and thus probing the details
of level repulsion. The effective model of a triangle of Heisenberg spins
and weak anisotropies, as has been used to  model the molecular magnets $\{V_{15}\}$ and $\{Cu_3\}$, is used to illustrate this.
\PACS{	{75.50.Xx}{Molecular magnets} \and
		{75.60.Ej}{Magnetization curves, hysteresis, Barkhausen and related effects}
 \and {74.25.Nf}{Response to electromagnetic fields (nuclear magnetic resonance, surface impedance, etc.)}
     } 
} 
\maketitle
\section{Introduction}
\label{intro}
Studies of the quantum behavior of open systems are almost as old as quantum mechanics itself. The related phenomena of relaxation and decoherence have passed from conceptual to pratical questions with the improvements of experimental technique during the last decades. One can now  build and manipulate physical objects at the borderline between quantum and classical worlds. 
In the future, materials structured on such a scale may be of use in applications such as very tiny components of magnetic memory and quantum computers. This would require detailed understanding of switching and dissipation between different quantum states. For such memories we can distinguish ``dynamical'' aspects, related to writing (or reading) of different states, and ``relaxational'' aspects, concerning the storage itself. If such memories were to be part of future quantum computers, the phase of the wavefunctions is essential and the loss of phase central to the questions of the feasibility of building such a device. It is therefore crucial to understand and control the coupling of the small magnetic object to its environment.\par
There is a large literature on the subject of two level systems  in contact with an environment \cite{Leggett} but in the magnetic systems we shall consider, there are aspects not present in, for example, ``qubits'' based on Josephson couplings. In particular there is the complexity of the magnetic spectrum: as we shall see, it is essential to take into account more than  two levels to understand the dynamics. Furthermore the nature of the environment is unclear and new experiments should be designed that may shed light on the issue. This paper will address a  multi-level problem, calculating  the response of a relatively simple magnetic system to a fast-varying magnetic field, including dissipation to a  a simple  model  environment.  In the context of quantum computing this corresponds to the ``writing'' aspect on a potential quantum memory storage device. This will be illustrated using the example of a triangle of magnetic ions, as a model of low spin molecular magnets such as \{V$_{15}$\} or \{Cu$_3$\}. We choose the triangular geometry because of recent experiments, to be noted later, and because the spectrum is sufficiently rich that it displays at least some of the features not seen in simple two-level systems.
\par Experimentally, the development of new magnetic coils allowing for the generation of high magnetic fields varying on very short time scales has lead to  experiments that can probe such interesting aspects of  statistical mechanics. We shall argue that {\it non-equilibrium} measurements with pulsed magnetic fields can be more informative, if properly analysed, than equilibrium measurements for  fine details of interactions between the spins. 
Such experiments usually involve  solids, where the molecular units repeat, and are assumed to have sufficiently weak interactions that they may be considered as independent, to a first approximation. They are therefore analysed in terms of a static Hamiltonian for the individual molecule, with an additional  time-dependent external field. The dissipation is induced by an environment, which represents contributions from coupling to phonons, dipole interactions with other molecules, hyperfine coupling to nuclear spins.

\section{Molecular magnets and Landau-Zener regime}
If the  interactions between spins were perfectly isotropic, transitions in magnetization would occur by the  crossing  of eigenvalues of different spins. In real magnets an\-isotropies mix the levels and, on some scale, crossings are avoided. The dynamics will therefore depend strongly on the fine structure of the molecular levels of the spins. If the level repulsion is weak, as is the case if the anisotropies are small, it may be possible to enter a r\'egime where the temperature is sufficiently low, and the coupling to the environment sufficiently weak, that we can be in a regime of pure quantum tunneling. It is for this reason that there has been great interest in the Landau-Zener tunelling r\'egime~\cite{Mn12}.\par
If we consider {\it only two} levels with an avoided crossing, the transitions between the two levels, when varying a control parameter such as the external magnetic field, is described by the Landau-Zener formula \cite{LZ}:
\bb
\mathcal{P}=1-\exp\left( -\frac{\pi \Delta^2}{\hbar 2 v}\right),
\ee
where the ``velocity'' $v$ is defined by $v=\frac {\dd(E_1-E_2)}{\dd t}$, and $E_1$ and $E_2$ are the eigenvalues of the two levels considered. $\mathcal{P}$ is the asymptotic probability for passing to the upper level given the initial condition of a state in the lower. We note that this formula is only for the asymptotic probability of transfer and is for zero temperature; thus it does not fully determine  the dynamics. It does, however, suffice to see when quantum tunneling can be relevant. For the argument of the exponential to be sufficiently small that the probability of staying in the lowest level is other than unity, the ``velocity'' $v$ must usually be extremely large, unless the gap is very tiny: e.g. for $\Delta = 1 \mathrm{K}$, quantum tunneling is relevant for velocities greater than $v \approx 10^{11} \mathrm{T s}^{-1}$. Thus to attain this regime either $\Delta$ must be extremely tiny or $v$ very large. Two recent sets of experiments do, however, approach the tunneling regime:
\par (i) In the case of high spin magnets such as \{Mn$_{12}$\} \cite{Mn12} or \{Fe$_8$\} \cite{WernsdorferFe12}, the two nearly degenerate states correspond to two states of total magnetization $S_z=\pm 10$  and tunnelling is  via a series of virtual transitions to states of smaller $| S_z |$ in the multiplet of total $S=10$. The  energies  of these states are determined  primarily by the easy axis anisotropy $DS_z^2$.\cite{Politietal} Since the splitting is tiny, the tunneling probability can be appreciable for very modest changes in  the  external magnetic field. The experiments are, of course, carried out at temperatures much greater than the energy splitting. 
\par (ii) In  low spin magnets such as \{V$_{15}$\} \cite{V15} or \{Cu$_{3}$\} \cite{Choi},  the active degrees of freedom can be 
taken to be essentially coupled spin 1/2 moments, seated on the vertex of equilateral triangles.
In \{V$_{15}$\} these may be thought of as the three spin 1/2 V$^{4+}$ ions sandwiched between two planes of six ions. As de Raedt {\it et al} have shown\cite{deRaedtMiyashita}, the extra Vanadium ions renormalize the effective interactions in the plane, but for the fields of interest the out-of-plane spins can otherwise be considered as being locked in inert singlet states. The gaps are generated by Dzalyoshinsky-Moriya (DM) interactions and are typically $10^{-2}$ K so that the velocity needed for Landau-Zener tunnelling would be 10$^{7}\mathrm{ T.s^{-1}}$, within range of recent ``single-turn'' magnets \cite{RapidFields}. The advantage of these compounds compared to those of (i) is that experiments can be conducted at temperatures comparable to the splittings, thus a different r\'egime of non-equilibrium phenomena can be studied. Furthermore Choi \textit{et al.} have suggested that the extra degeneracy of different chiral states of the triangular structures may be useful for applications \cite{Choi}.\par
A problem in the interpretation of past experiments is that it is not necessarily clear  whether one is  really in the tunneling regime. One approach has to be to look for the characteristic dependence on the sweeping velocity of the magnetic field and compare to the predictions of Landau-Zener theory. The difficulty is that features found such as the magnetic Foehn effect \cite{Foehn}, where extra plateaux in the magnetization occur in the non-equilibrium response, may be described either as phonon bottleneck effects \cite{Adiabatic2,Butterfly,Adiabatic}, i.e. that the bath does not relax to equilibrium on the timescale of the experiment, or by a dynamical effect of the system coupled to a bath that stays in equilibrium. In what follows we shall look for extra features which could distinguish effects intrinsic to the small quantum system from such  bottleneck effects. \par

\begin{figure}[tbp]
\subfigure[Energy spectrum as a function of the magnetic field. Both energies and fields are in units
of J.]{\includegraphics[width=\linewidth]{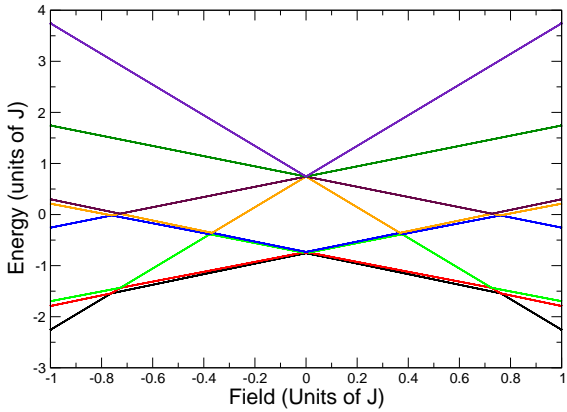}}
\subfigure[Equilibrium magnetization (in units of $\mu_B$) in the direction of the field for $T=0.1J$.]{\includegraphics[width=\linewidth]{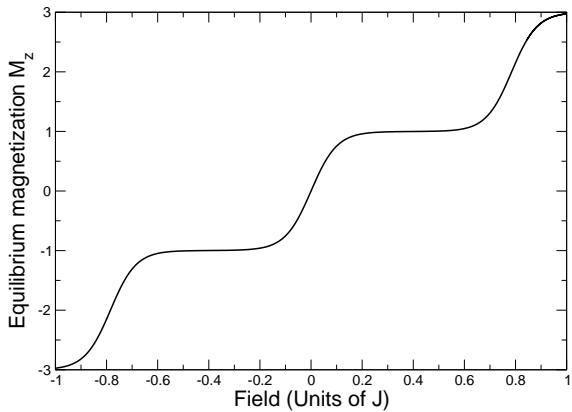}}
\caption{(Color online) Spectrum of Hamiltonian ${\mathcal H}_0$
and equilibrium magnetization.}
\label{spectrum}
\end{figure}

\section{Model for the molecular magnets \{V$_{\mathsf{15}}$\} or \{Cu$_\mathsf{3}$\}}
In order to describe the reaction of the magnet to the external magnetic field, we have to precise the internal Hamiltonian for the molecular magnets. For systems (ii), to which the rest of the paper is devoted, we can write the Hamiltonian for three  spin $\frac 1 2$ moments 
\bb \nn
\mathcal{H}_0(t)=\sum_{i=1}^3\left(J_{i}\vec{S}_i \cdot \vec{S}_{i+1}+\vec{D}_{i}\cdot\left(\vec{S}_i \times \vec{S}_{i+1}\right)+\vec{H}(t)\cdot \vec{S}^z_i\right).
\ee
The system has  $2^3=8$ energy levels. The exchanges are chosen for triangular symmetry, i.e. the isotropic exchanges are equal and the three vectors defining DM interactions are determined by a single vector $\vec{D}$ and rotated around the three-fold  symmetry axis \cite{deRaedtMiyashita}. Before discussing the dynamics, we shall discuss the static properties of ${\cal H}_{0}$. In Figure \ref{spectrum} we show its  spectrum , i.e. with $\vec{H}(t)=H_z \vec{e}_z$ constant for  different values of $H_z$.   At large fields the spin $\frac{3}{2}$ components have lowest energies and cross with the two degenerate spin  $\frac{1}{2}$ levels of opposite chirality. To make the degeneracies apparent a small chiral-breaking term has been added. We choose this additional term to be a Sen-Chitra interaction $\mathcal{H}_\mathrm{chiral}={\mathrm{A_{chiral}}}\left(\vec{H} \cdot \vec{e}_z\right) \;\vec{S}_1 \cdot \left(\vec{S}_2 \times \vec{S}_3\right)$ \cite{Chitra}. The small Dzalyoshinsky-Moriya interactions give avoided level crossings. Below the spectrum we show the equilibrium magnetization at temperature $\kb T=0.1J$. It is clear that unless extremely low temperatures are available, little information on the details of the level crossing can be found, other than the points where levels cross, i.e. the value of the isotropic  exchange.\par
In contrast to the relatively featureless equilibrium magnetization it will be seen that {\it non-equilibrium} observables such as magnetization are much richer  and we now turn to the question of dynamics. We remark also that in such triangular-based spin systems, switching between different states involves more than just two states at a time; because of the chirality the crossings involve three (between spin $\frac{3}{2}$ and  the two chiral $\frac{1}{2}$ states) or four (between the degenerate chiral doublet at $\vec{H}=0$). At the least a generalization of the Landau-Zener form is required. We remark that in certain cases the asymptotic transition probabilities for such  multi-crossing may be reduced, via an ``Independent Crossing Approximation''  \cite{ICA} to the calculation of two-level crossings. Because of the importance of avoided crossings we need a general theory that applies for multiple crossings with the details of
the anisotropy \cite{Foldi}. For direct comparison with experiment it is also useful to have  a general form of the time-dependent external field generated by the coils. \par

\section{Environment: model and coupling to the magnet}
In current experiments and probably future applications, the magnet can not be considered as a isolated system. It is indeed surrounded by an environment, responsible for relaxation and decoherence phenomena. 
In order to take into account those effects in our calculation, we have to 
define an  environment explicitly. Unfortunately 
the true environment is not yet known precisely. We 
therefore focus here on a simple  environment \cite{MaximeInPrep} modelled as an infinite set of bosonic harmonic oscillators:
\bb
\mathcal{H}_\mathrm{env}=\int_{-\infty}^{+\infty} \dd \omega\; \omega I(\omega) \left( b^{\dagger}_\omega b_\omega+\frac{1}{2}\right),
\ee
We choose the spectral density $I(\omega)$ to be a power law,
\bb
I(\omega)=\omega^\alpha \Theta(\omega),
\ee
with $\alpha=2$ (Super-Ohmic regime, as appropriate for a three dimensional phonon bath), and where $\Theta$ is the Heaviside distribution. Several additional hypotheses will be needed in the following. First we will supposed that the initial state of environment is a thermal state, that is to say that the environment is in thermal equilibirum with a thermostat at a given temperature $T$. Furthermore the environment is assumed to be Markovian, i.e.  its correlation time  is much shorter than the  time scale of the dynamics. Taking these two hypotheses together  implies a static  environment, described at all times by the density matrix $\rho_\mathrm{env}(t)\propto \exp\left(-\beta H_\mathrm{env}\right)$ \cite{Blum,Breuer}.\\
The coupling of the spin system to the bath is included with the term 
\bb
\mathcal{H}_\mathrm{coupling}=g \int_{-\infty}^{+\infty} \dd \omega\; I(\omega) (b^{\dagger}_\omega+b_\omega)\vec{X} 
\ee
The operator $\vec{X}$  is some combination of spin operators $\mathbf{S}_i$ and  couples the spin system linearly to the boson operators of the bath, $b_\omega$ $b^\dag_\omega$, with dimensionless strength $g$.  It has sufficiently low symmetry that it can restore equilibrium from any initial state. e.g. $\vec{X}=\sum_{i=1}^3 (a_i\vec{S}^x_i+b_i\vec{S}^y_i+c_i\vec{S}^z_i)$. The  coefficients $a_i,b_i,c_i$  differ from spin to spin, in order to induce transitions
between states of different chiralities. 

\section{Bloch-Redfield approach to the reduced density matrix dynamics}
The system (magnet + environment) is a closed quantum system described by the Hamiltonian
\bb
\mathcal{H}_\mathrm{tot}=\mathcal{H}_0(t)+\mathcal{H}_\mathrm{env}+\mathcal{H}_\mathrm{coupling}.
\ee
All the physical information of this {\it closed}  system can be determined by  the density matrix $\rho_{\mathrm{tot}}(t)$. As the experimental set-up allows measurements on the magnet only, and not on the environment, we are interested in this magnet as an {\it open} quantum system, described by the reduced density matrix $\rho(t)=\mathrm{Tr}_\mathrm{env} \left[ \rho_{\mathrm{tot}}(t) \right]$. A na\"\i ve approach to study the magnetic dynamics would be to integrate the Liouville equation, coming from  first principles, to obtain the  total density matrix $\rho_{\mathrm{tot}}(t)$ at all time, and then to perform the trace on the environment degrees of freedom. However such an approach is essentially impossible to follow in practice given the high dimension of the Hilbert space. An approach to this problem, that is  by now standard, is to make the additional hypotheses on the environment presented in the previous section, to obtain a dynamical master equation for the reduced density matrix. This  perturbative approach dates back to the 1950's \cite{Bloch,Redfield,Kubo}. We would like to stress that the equation does not come directly from  first principles: it relies on additional assumptions whose validity needs to be tested for each particular system. For systems with a larger number of levels, the equations have been put in a compact form \cite{deRaedtMiyashita,Miyashita} to give the equation for the time dependence (in units where $\hbar=1$ )of the density matrix projected onto the Hilbert space $\rho(t)$ of the molecular system (in our case the three coupled spins):
\bb \label{qme}
\frac{\partial \rho}{\partial t}=-{i}[\mathcal{H}_0(t),\rho(t)]-g \Big([\vec{X},R\rho(t)]+ [\vec{X},R\rho(t)]^\dagger \Big).
\ee
If we define the instantaneous eigenbasis \{$|k,t\rangle$\} of $\mathcal{H}_0(t)$ by
\bb
H_0(t) |k,t\rangle = E_k(t) |k,t\rangle,
\ee
the operator $R(t)$ is determined by its matrix elements  
\bb
\langle k,t |R| m,t \rangle=\Big[I\big(\omega_{km}(t)\big)-I\big(-\omega_{km}(t)\big)\big]\\ \nn \times n_\beta\left[\omega_{km}(t)\right]X_{km}(t),
\ee
where $\omega_{km}(t)=E_k(t)-E_m(t)$ and
\bb
n_\beta(\omega)=(e^{\beta \omega}-1)^{-1}.
\ee
The external field $\vec{H}$ is taken to have arbitrary time variation in a single direction $\vec{e}_x$, $\vec{e}_y$ or $\vec{e}_z$. \\

\begin{figure}[tbp]
\includegraphics[width=\linewidth]{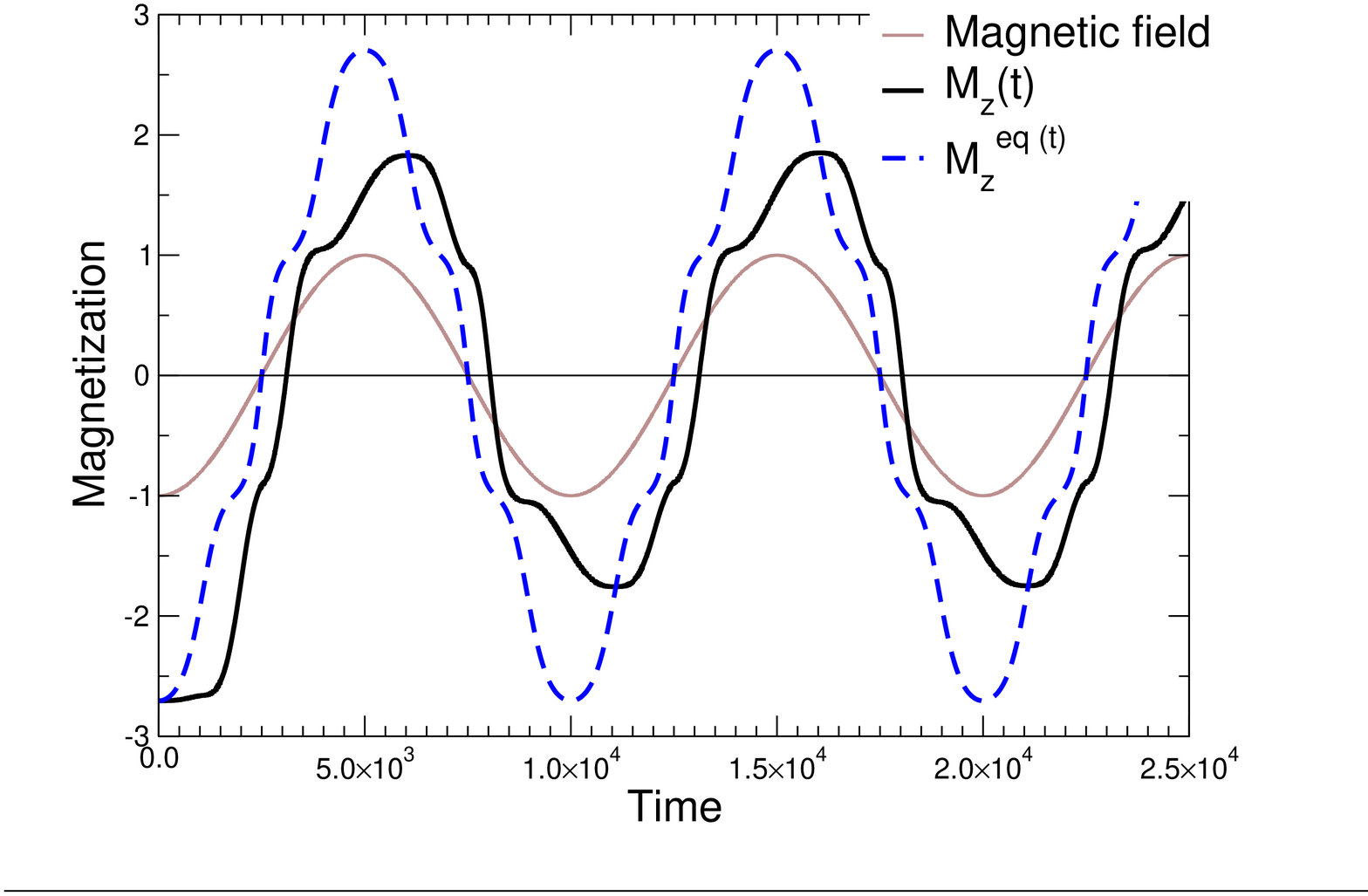}
\caption{(Color online)  Magnetization as a function of time. The applied field is  $\vec{H}(t)=H_z(t)\vec{e}_z$ (\ref{field}). $H_z(t)=-H_0\cos(\Omega t)$. The blue lines shows the equilibrium magnetization for each instantaneous value of the field. The parameters used are the following:  $H_0=J$, $\Omega=2\pi \times 10^{-4} \mathrm{J} / {\hbar} $, $D_z=0.05\mathrm{J}$, ${\mathrm{A_{chiral}}}=J/20 $ and $T=0.2 J$. The coupling $g=4\times 10^{-3}$} 
\label{mlongt}
\end{figure}
We shall now present results based on integration of the quantum master equation (\ref{qme}). In practice we integrate by fourth order Runge-Kutta approximation which leads to numerically convergent
results, at least for weak coupling $g$ and for times long enough to define several hysteresis loops. 
The hamiltonian must be diagonalised for each time step in order to define the matrix $R(t)$. The time variation
of the external field can be arbitrary but for our purposes we shall take it as cosinusoidal:
\bb
\vec{H}(t)=H_0\cos(\Omega t)\vec{e}_i,\; i=x,y,z,
\label{field}
\ee
and as initial conditions $\rho(t=0)=\exp(-\beta{\mathcal{H}}_0(t=0))$, i.e. the spin system in equilibrium with the heat bath at inverse temperature $\beta=\frac 1 {\kb T}$.\\

\section{Numerical observations}
We shall now make calculations of the observables for a model case in which we take a single component of the DM vector along the $z$-direction, $\vec{D}=D_z \vec{e}_z$. The purpose here is not to fit a particular experimental system, more to make general statements. We first show the three components of magnetization as a function of time. The initial condition is the equilibrium density matrix for the same temperature and the field $H(t=0)=-J$. It is seen that at the relatively weak couplings that can be integrated numerically ($g=10^{-3}$ to $10^{-4}$) the magnetization ``lags'' the equilibrium magnetization. After one full oscillation the profile (in time) of the out-of-equilibrium magnetization approximately repeats, converging slowly towards  some steady-state form. We remark that the period for the sinusoidal external field is $10^4$ in units of $J^{-1}$. Returning to standard units, where $\hbar$ is not unity, this corresponds to a time of $5\times 10^{-9}$ seconds for J=1 K.\par
\begin{figure*}[tbp]
\subfigure[Hysteresis loops $M_z(H_z)$.\label{looppar}]{\includegraphics[width=0.45\linewidth]{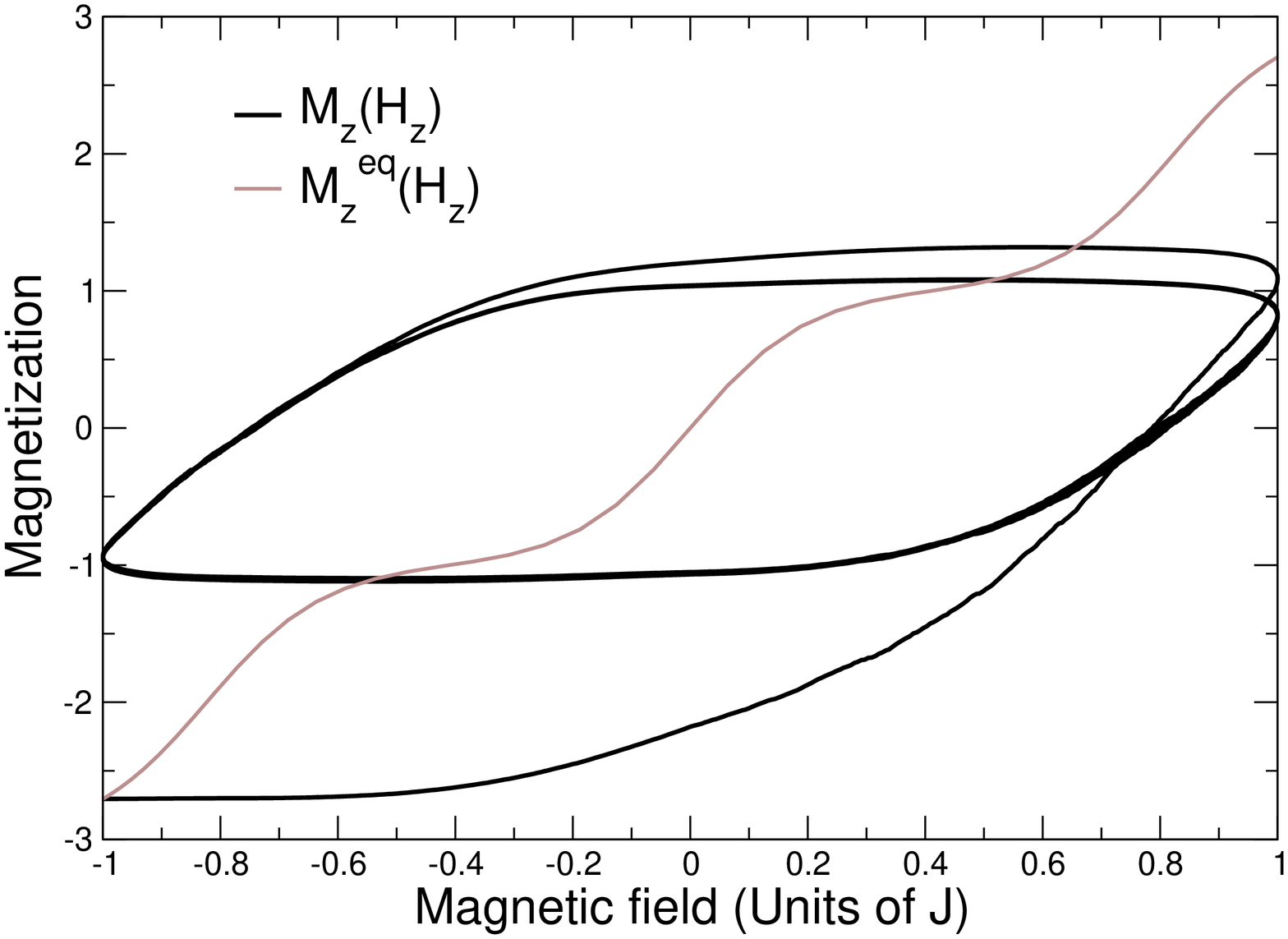}}\hfill
\subfigure[Detail of the spectrum. \label{zoomspectpar}]{\includegraphics[width=0.45\linewidth]{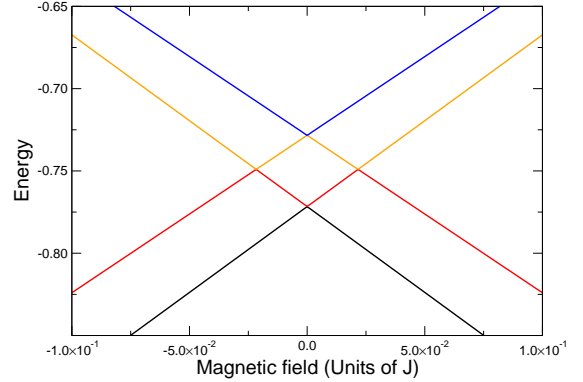}} 
\caption{(Color online) Energy spectrum and dynamics of the magnetization parallel to the applied field, for $\vec{H}=H_z(t)\vec{e}_z$ parallel to  Dzalyoshinsky-Moriya vector $\vec{D}=D_z \vec{e}_z$. The parameters are $D_z=J/20$, $H_0=J$, $\Omega=2\pi \times 10^{-3} \mathrm{J} /{\hbar}$, $g=4\times 10^{-3}$ and $T=0.2J$. In the righthand panel, the details
of the spectrum of the four lowest energy levels around zero field of Fig. \ref{spectrum}, showing the crossing of the two nearly degenerate doublets.}
\label{dmloops}
\end{figure*}
\begin{figure*}[tbp]
\subfigure[Hysteresis loops $M_x(H_x)$.\label{loopperp}]{\includegraphics[width=0.45\linewidth]{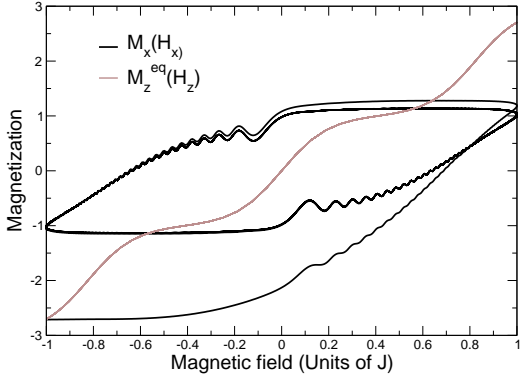}} \hfill
\subfigure[Detail of the spectrum.\label{zoomspectperp}]{\includegraphics[width=0.45\linewidth]{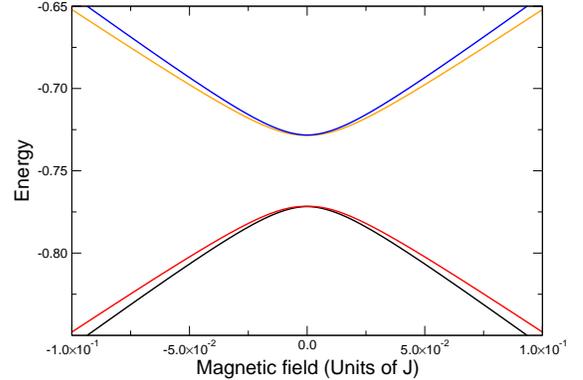}} 
\caption{(Color online) Energy spectrum and dynamics of the magnetization parallel to the applied field, for $\vec{H}=H_x(t)\vec{e}_x$ perpendicular to  Dzalyoshinsky-Moriya vector $\vec{D}=D_z \vec{e}_z$. The parameters  are identical
to those of Figure \ref{dmloops}: $D_z=J/20$, $H_0=J$, $\Omega=2\pi \times 10^{-3}\frac {\mathrm{J}}{\hbar}$, $g=4\times 10^{-3}$ and $T=0.2J$. In the right hand panel the avoided level crossing of the doublet states at zero field.}
\label{crossingzero}
\end{figure*}

The information in Figure \ref{mlongt} can be represented in a form  often used by experimentalists, i.e. as a hysteresis loop, where the magnetization is plotted against the external field. This we do in Figures \ref{looppar} and \ref{loopperp}.  The difference between the two hysteresis loops  is that in Figure \ref{looppar}, the DM vector is parallel to the external field direction and in the second, Figure \ref{loopperp}, it is perpendicular to the DM vector. In each case the magnetization is longitudinal, i.e. parallel to the applied field.
The differences between the two orientations $m_z\left(H_z\right)$ and $m_x\left(H_x\right)$ are that the hysteresis loop is ``fatter'' when the field is perpendicular to the DM vector, but, more strikingly, that there are extra oscillations visible in $m_x\left(H_x\right)$ that are absent for $m_z\left(H_z\right)$. What is the difference between the two field orientations in terms of the spectrum of excitations? If we look at an enlargement of the spectrum of Figure \ref{spectrum} around zero field, as shown in Figures \ref{zoomspectpar} and \ref{zoomspectperp} this difference is apparent. For $m_x\left(H_x\right)$ there is an avoided level crossing at zero field with a minimum splitting whereas for $m_z\left(H_z\right)$ there is a crossing of levels. This avoided level crossing makes the hysteresis loop ``thinner'' as the lowest level smoothly evolves from one spin state to another.
\begin{figure}[tbp]
\includegraphics[width=\linewidth]{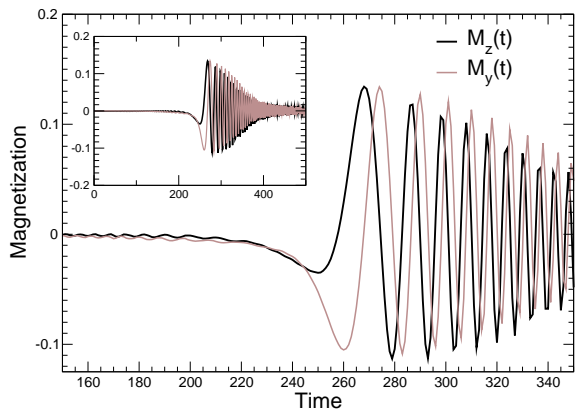}
\caption{(Color online) Details of the oscillations in the magnetisation transverse to the applied field, perpendicular to the DM vector. Inset: Wave packet on a longer time scale. The parameters are $D_z=J/20$, $H_0=J$, $\Omega=2\pi \times 10^{-3}\mathrm{J}/{\hbar}$, $g=8\times 10^{-3}$ and $T=0.5J$. The avoided crossing is centred at t=250.}
\label{transverseblowup}
\end{figure}
So what is the significance of the oscillations? For a two-level system formulated as a (pseudo-) spin in transverse field and time-varying longitudinal field, these transient oscillations\cite{Miyashita} correspond to precession around an instantaneous axis which, near the avoided crossing, is perpendicular to the applied field. In other words these are simply Rabi oscillations in the presence of a time-varying gap. These oscillations are not periodic, but the characteristic frequency range is the gap in the avoided level crossing. This mixing of longitudinal and transverse oscillation is well known in ``zero-field'' NMR~\cite{NMR}. This suggests that the oscillations may be more visible in the transverse components of magnetization, i.e. the components perpendicular to the applied field. In Figure \ref{transverseblowup} we show the transverse magnetizations, for applied field perpendicular to the DM vector. It is seen that they are small except close to avoided crossings, where there are oscillations rapid on the time scale of the driving field. We remark that they are visible at a temperature (0.5J) well 
above the gaps induced by anisotropies.\par
This leads to the suggestion for experiment that instead of looking exclusively at the shape of the hysteresis loop, as has been done most in the past; it may be more instructive to look at the Rabi oscillations. They may be most visible  in the transverse components,  as they are against a flat background. We remark that these oscillating components have  been seen clearly in past numerical calculations \cite{Miyashita,Foehn,Rau} of the longitudinal magnetization but have attracted relatively little attention. We recall however from the theory  of two-level systems that the decay of longitudinal components and transverse oscillations at a fixed magnetic field determines two characteristic time scales  which give different information on the spectrum of the bath: in particular the transverse oscillations are sensitive to the zero frequency component of the bath \cite{T1T2}. By analogy, measuring both longitudinal and transverse components one should, in principle at least, determine different properties of the  spectrum of the bath.\par
We remark that for quantitative studies of the oscillations we should be  a little more precise about what we mean by ``transverse''; ideally we would take a component with vanishing mean. For the simple case we have shown  here the ideal component is that which is perpendicular to both the applied field (parallel to $\vec{e}_x$) and parallel to the DM vectors, i.e. $m_z\left(H_x\right)$. The component perpendicular to both the field and the  DM vector $m_y\left(H_x\right)$, has small but non-zero values depending on the field.  For non-collinear DM vectors on the three spins there may be no absolute separation of smooth and oscillating contributions; nevertheless the oscillations will always be stronger perpendicular rather than parallel to the applied field.\par
We note that these oscillations  occur when the rotational symmetry is broken. This  occurs if, as is the case in Figure \ref{transverseblowup}, the DM vector and the external field completely break rotational symmetry, so that the phase transverse is determined by the (time-dependent) Hamiltonian \cite{NojiriRemark}. Oscillations  also occur, (but are not shown in the Figures), in   much smaller amplitudes of oscillations for $m_x\left(H_z\right)$; that is even when the Hamiltonian ${\cal{H}}_0$ is symmetric around the field direction.  This can happen because the coupling to the bath breaks the symmetry in our simulations. Whether this is true in a physical system or not, depends on the symmetries  of the environment of the magnetic molecule. To restore symmetry in our formulation, while nonetheless allowing relaxation,  we could include a sum of several terms of the form written for ${\cal{H}}_\mathrm{coupling}$.

\section{Conclusions}
We have  calculated the dynamic behaviour of weakly an\-isotropic triangular antiferromagnets, as could be seen in the molecular magnetis \{V$_{15}$\} or \{Cu$_{3}$\} subject to vary fast varying magnetic fields.  Our simulations  model in  detail  effects of tunneling including transient effects in a situation where Landau-Zener expressions are not sufficient to determine dynamics at arbitrary frequency and at finite temperature. This is  especially so here are several levels contribute to the response because of degenerate  chiral states on the triangle. We have argued that a useful characterization of the tunneling r\'egime, useful would be the  (non-periodic) oscillations in the magnetizations and, in particular, in the transverse components close to regions of level anti-crossing. This should yield more detailed information on the anisotropies causing level repulsion in molecular magnets, and the coupling to the bath, e.g.  the value of the coupling $g$ and the nature of the bath. The oscillations can persist (at least at weak coupling) up to temperatures well above the scale of the avoided level crossing. To illustrate this we have calculated  hysteresis loops for a model in which there is a single component to the DM vector and the field is taken either parallel or perpendicular to this vector. We  chose the  simple case of a  single component  DM vector in order to illustrate the phenomenon. For this case there is only a single region of avoided crossing at zero field. In real systems such as \{Cu$_{3}$\} all three components are  present. There are then avoided crossings where, in the isotropic limit, the spin $\frac 3 2$ cross
the spin $\frac 1 2$ levels $H = \pm {\frac {3}{4}} J$, and  oscillations will occur at all crossings.
\par We conclude that if one is interested in studying the physics  at the scale of a single molecular magnet, it might be useful to study not only the traditional hysteresis loops, but also the transient oscillations, visible in the transverse component of the magnetization dynamics, as the first correction to the adiabatic limit.


\begin{thebibliography}{}
\bibitem{Leggett} A.O. Caldeira and A.J. Leggett, Annals of Physics \textbf{149} (2): 374-456 (1983); Y. Makhlin \textit{et al.} Rev. Mod. Phys. \textbf{73}, 357 (2001).
\bibitem{Mn12}L. Thomas, F. Lionti, R. Ballou, D. Gattesehi, R. Sessoli, and B. Barbara, Nature (London) (London) {\bf 383}, 145 (1996).
\bibitem{LZ}L. Landau, Phys. Z. Sowjetunion {\bf 2}, 46 (1932);
C. Zener, Proc. R. Soc. London A {\bf 137}, 696 (1932);
E. C. G. St\" uckelberg, Helv. Phys. Acta {\bf 5}, 369 (1932).
E. Majorana, Nuovo Cimento {\bf 9} 43(1932).
\bibitem{WernsdorferFe12} W. Wernsdorfer,   R. Sessoli Science {\bf 284} 5411 (1999).
\bibitem{Politietal} P. Politi A. Rettori, F. Hartmann-Boutron, J. Villain Phys. Rev. Lett. {\bf 75}, 537 (1995).
\bibitem{V15}Chiorescu, I ,Wernsdorfer, W.,  M\"uller, A., B\"ogge, H., Barbara, B. , Phys. Rev. Lett. {\bf 84}3454 (2000).
\bibitem{Choi}K.-Y. Choi, Y. H. Matsuda, H. Nojiri, U. Kortz, F. Hussain, A. C. Stowe, C. Ramsey, and N. S. Dalal, Phys. Rev. Lett.{\bf  96}, 107202 (2006).
\bibitem{deRaedtMiyashita} H. De Raedt H, S. Miyashita, K.Michielsen, M. Machida, Phys. Rev. B {\bf 70} 064401 (2004).
\bibitem{RapidFields} H. Nojiri \textit{et al.}, Physica {\bf B} 346-347 (2004) 216, A. Kirste, Doctoral thesis, Humboldt University, Berlin ( 2004).
\bibitem{Foehn} K. Saito and S. Miyashita, J. Phys. Soc. Jpn. {\bf  70}, 3385 (2001).
\bibitem{Adiabatic2} I. Rousochatzakis and M. Luban, Phys. Rev. B {\bf 72}, 134424 (2005).
\bibitem{Butterfly}M. Vogelsberger and D. A. Garanin, Phys. Rev. B {\bf 73}, 092412 (2006).
\bibitem{Adiabatic}I. Rousochatzakis, Y. Ajiro, H. Mitamura, P. Kerler, and M. Luban, Phys. Rev. Lett. {\bf 94}, 147204 (2005).
\bibitem{Chitra} D. Sen and R. Chitra, Phys. Rev. B \textbf{51}, 1922 (1995).
\bibitem{ICA} Y.N. Demkov and V.N. Ostrovsky, J. Phys. B {\bf 34}, 2419 (2001);
S. Brundobler and V. Elser J.Phys. A: Math. Gen {\bf 26}, 1211 (1993).
\bibitem{Foldi} P. F\"oldi, M. G. Benedict,J.Milton Pereira Jr. F. M. Peeters, Phys. Rev.  B {\bf 75}, 104430 (2007) have also emphasised the need need to go beyond Landau-Zener theory for the calculation of asymptotic
transition probabilities.
\bibitem{MaximeInPrep} Further discussions on the nature of the environment and the adiabatic limit will be presented in a forthcoming article.
\bibitem{Blum} K. Blum, \textit{Density Matrix Theory and Applications}, Plenum Publishing Corporation, 2nd edition (1996).
\bibitem{Breuer} H.-P. Breuer and  F. Pettrucionne, \textit{The Theory of Quantum Open Systems}, Oxford University Press (2002).
\bibitem{Bloch} F. Bloch, Phys. Rev. {\bf 70}, 460 (1946);
{\it ibid} {\bf 102}, 104 (1956); 
{\bf 105}, 1206 (1957).
\bibitem{Redfield}A. G. Redfield Phys. Rev. {\bf 98} 1787 (1955).
\bibitem{Kubo} R. Kubo,M. Toda and N. Hashitsume {\it Statistical Physics II} (Spinger-Verlag, N.Y. 1985).
\bibitem{NMR}R. Kubo and K. Tomita, J. Phys. Soc. Jpn. \textbf{9}, 888 (1954). 
\bibitem{Miyashita} S. Miyashita, J. Phys. Soc. Jpn. {\bf 64}, 3207 (1995). 
\bibitem{Rau}
A.R.P. Rau and R.A. Wendell, Phys. Rev. Lett, \textbf{89}, 220405, (2002).
\bibitem{T1T2} Cohen-Tannoudji C \textit{et al.}, \textit{Atom-Photon interactions}, Wiley (1988).
\bibitem{NojiriRemark} We thank H. Nojiri for raising this issue.
\end{thebibliography}
\end{document}